\documentclass[english,aps, prb, superscriptaddress, twocolumn,a4paper,nofootinbib]{revtex4}
\usepackage[T1]{fontenc}
\usepackage[latin9]{inputenc}
\usepackage{geometry}
\usepackage{color}
\usepackage{babel}
\usepackage{array}
\usepackage{booktabs}
\usepackage{amsmath}
\usepackage{graphicx}
\usepackage[unicode=true,pdfusetitle,
 bookmarks=true,bookmarksnumbered=false,bookmarksopen=false,
 breaklinks=true,pdfborder={0 0 0},pdfborderstyle={},backref=false,colorlinks=true]
 {hyperref}

\makeatletter

\providecommand{\tabularnewline}{\\}

\@ifundefined{textcolor}{}
{%
 \definecolor{BLACK}{gray}{0}
 \definecolor{WHITE}{gray}{1}
 \definecolor{RED}{rgb}{1,0,0}
 \definecolor{GREEN}{rgb}{0,1,0}
 \definecolor{BLUE}{rgb}{0,0,1}
 \definecolor{CYAN}{cmyk}{1,0,0,0}
 \definecolor{MAGENTA}{cmyk}{0,1,0,0}
 \definecolor{YELLOW}{cmyk}{0,0,1,0}
}

\usepackage{booktabs}

\makeatother

\begin{document}

\title{Mott physics and collective modes: an atomic approximation of the
four-particle irreducible functional}

\author{Thomas Ayral}
\email{thomas.ayral@cea.fr}

\selectlanguage{english}%

\affiliation{Institut de Physique Théorique (IPhT), CEA, CNRS, UMR 3681, 91191
Gif-sur-Yvette, France}

\author{Olivier Parcollet}

\affiliation{Institut de Physique Théorique (IPhT), CEA, CNRS, UMR 3681, 91191
Gif-sur-Yvette, France}
\begin{abstract}
We discuss a generalization of the dynamical mean field theory (DMFT)
for strongly correlated systems close to a Mott transition based on
a systematic approximation of the fully irreducible four-point vertex.
It is an atomic-limit approximation of a functional of the one- and
two-particle Green functions, built with the second Legendre transform
of the free energy with respect to the two-particle Green function.
This functional is represented diagrammatically by four-particle irreducible
(4PI) diagrams. Like the dynamical vertex approximation (D$\Gamma$A),
the fully irreducible vertex is computed from a quantum impurity model
whose bath is self-consistently determined by solving the parquet
equations. However, in contrast with D$\Gamma$A and DMFT, the interaction
term of the impurity model is also self-consistently determined. The
method interpolates between the parquet approximation at weak coupling
and the atomic limit, where it is exact. It is applicable to systems
with short-range and long-range interactions.
\end{abstract}
\maketitle

\section{Introduction}

Strongly correlated electron systems pose a great challenge to theoretical
physics. Not only is the direct solution of strongly interacting lattice
models thwarted by the exponential size of the Hilbert space (or the
corresponding negative sign problem in quantum Monte-Carlo simulations),
but it is also difficult to find controlled approximate methods in
regimes of physical interest.

One such class of methods is dynamical mean field theory\cite{Georges1996}
(DMFT) and its cluster extensions\cite{Lichtenstein2000,Kotliar2001,Hettler1998,Hettler1999,Maier2005a},
which are based on an expansion around the atomic limit of the two
particle irreducible (2PI) or Luttinger-Ward functional $\Phi_{\mathrm{LW}}$.
This local expansion is performed by mapping the extended lattice
problem onto an effective impurity problem with the same interaction
vertex as the lattice's and a dynamical bath describing the incursions
of electrons on and off the impurity. The self-energy of the impurity
is used to approximate the lattice self-energy. Despite a number of
successes in describing important features of strongly correlated
electron systems, cluster DMFT methods are limited by the maximal
cluster size attainable by quantum Monte-Carlo solvers, and cannot
describe the effect of collective modes with a range exceeding the
size of the cluster.

The dynamical vertex approximation\cite{Toschi2007,Katanin2009,Held2014}
(D$\Gamma$A) proposes to approximate not the self-energy, but the
four-leg vertex function by its impurity counterpart.\cite{Rohringer2012,Schafer2013}
This approximation is based on numerical hints that the fully irreducible
vertex is more local in space than the self-energy\cite{Maier2006}
and on the premise that two-particle quantities not only have an important
feedback on one-particle observables, but are also essential to understand
the underlying physical processes.\cite{Gunnarsson,Gunnarsson2016}
In this method, the local vertex from a converged DMFT computation
\textendash{} whether the fully irreducible vertex $\Lambda_{\mathrm{imp}}$
(``parquet D$\Gamma$A''\cite{Valli2014,Li2015a}) or the irreducible
vertex in a given channel, $\Gamma_{\mathrm{imp}}^{r}$ (``ladder
D$\Gamma$A''\cite{Rohringer2011,Schafer2014,Schafer2015,Schafer2016};
see also Ref \onlinecite{Kusunose2006} for a simplified version of
this approximation) \textendash{} is used to compute the momentum-dependent
self-energy via the Schwinger-Dyson equation. In principle, this self-energy
can be used to update the bath $\mathcal{G}$ of the impurity model,
although this fully self-consistent version has thus far not been
implemented. A similar approximation of the irreducible vertex in
a given channel is introduced in Ref. \onlinecite{Slezak2009} and
solved self-consistently.

Contrary to DMFT, D$\Gamma$A has not been derived as the local approximation
of a functional. This paper intends to fill this gap. By giving a
functional footing to the local approximation of the vertex, we clarify
the links of D$\Gamma$A with the parquet formalism. Most importantly,
we obtain precise prescriptions to construct the impurity model in
a way that is consistent with the local vertex approximation. The
method we obtain is similar to D$\Gamma$A, with additional self-consistent
interactions that can take into account the feedback of collective
modes and/or long-range interactions onto the impurity model.

Functional routes to extend DMFT include extended DMFT (EDMFT\cite{Sengupta1995,Kajueter1996,Si1996})
and the recently introduced triply-irreducible local expansion (TRILEX)
method\cite{Ayral2015,Ayral2015c}. Both methods rely on the introduction
of auxiliary bosonic variables and the subsequent approximation of
the exact electron-boson $n$PI functional by an expansion around
the atomic limit, with $n=2$ for EDMFT and $n=3$ for TRILEX. The
explicit introduction of bosonic degrees of freedom allows for direct
insights into the influence of collective modes on fermionic observables.
Alternative methods include the GW+EDMFT method\cite{Biermann2003,Sun2008,Ayral2013},
which supplements the atomic expansion of the 2PI functional with
nonlocal diagrams, and the dual boson method\cite{Rubtsov2011,VanLoon2014,Stepanov2015},
which resorts to another type of auxiliary bosonic fields to describe
nonlocal fluctuations beyond DMFT and EDMFT.

In this paper, we generalize the TRILEX idea to the 4PI level, without
resorting to auxiliary bosonic fields. Starting from a problem with
quartic fermionic interactions, we propose to approximate the functional
$\mathcal{K}_{4}$, which is represented by all four-particle irreducible
diagrams, by an expansion around the atomic limit. We call this approximation
QUADRILEX (for QUADRuply-Irreducible Local EXpansion) to distinguish
it from D$\Gamma$A. Like D$\Gamma$A, this approximation entails
the locality of the fully irreducible vertex. However, contrary to
D$\Gamma$A, it gives different prescriptions on how to update the
action of the impurity model at the level of the interaction term.
The latter is renormalized in a self-consistent manner. This method
can be regarded as a straightforward extension of DMFT from the 2PI
to the 4PI level.

This paper is organized as follows: in Section \ref{sec:Derivation-of-the},
we derive the method using functionals. We then discuss the main implications
of this method and its relation to known approximations in Section
\ref{sec:Discussion}.

\section{Derivation of the Formalism\label{sec:Derivation-of-the}}

We focus on a generic electron-electron interaction problem defined,
in a path-integral formalism, by the following action:

\begin{equation}
S=-\bar{c}_{\bar{u}}G_{0,\bar{u}v}^{-1}c_{v}+\frac{1}{2}U_{v\bar{u}x\bar{w}}\bar{c}_{\bar{u}}c_{v}\bar{c}_{\bar{w}}c_{x}\label{eq:ee_action}
\end{equation}

Latin indices gather the Bravais lattice site index, imaginary time
and the spin index: $u\equiv(\mathbf{R}_{u},\tau_{u},\sigma_{u})$.
We denote outgoing (resp. ingoing) points by indices with (resp. without)
a bar. Einstein summation over repeated indices is implied, and $\sum_{u}$
stands for $\sum_{\mathbf{R}}\int_{0}^{\beta}\mathrm{d}\tau\sum_{\sigma}$.
$\bar{c}$ and $c$ are Grassmann fields. $G_{0,\bar{u}v}$ denotes
the free propagator of the fermions, while $U_{v\bar{u}x\bar{w}}$
is the four-fermion bare interaction vertex. This generic action encompasses
a number of well-known models for strongly-correlated systems such
as the Hubbard, extended Hubbard or $t$-$J$ models.

The partition function is defined as
\begin{equation}
Z[J,U]\equiv\int\mathcal{D}[\bar{cc}]e^{-S[U]+J_{\bar{u}v}\bar{c}_{\bar{u}}c_{v}}\label{eq:partition_function}
\end{equation}

where we have introduced a bilinear source term $J_{\bar{u}v}$. The
free energy is defined as: 
\begin{equation}
\Omega[J,U]\equiv-\ln Z[J,U]\label{eq:Omega_def}
\end{equation}

It is a functional of the bilinear source $J_{\bar{u}v}$ and of $U_{\bar{u}v\bar{w}x}$,
which can be regarded as a quadrilinear source. $\Omega[J,U]$ is
the generating functional of correlation functions. In particular,
the one- and two-particle Green's functions are given by:
\begin{align}
G_{u\bar{v}} & \equiv-\langle c_{u}\bar{c}_{\bar{v}}\rangle=-\frac{\partial\Omega}{\partial J_{\bar{v}u}}\Bigg|_{U}\label{eq:G_def_derivative}\\
G_{2,\bar{u}u\bar{v}v}^{\mathrm{nc}} & \equiv-\langle\bar{c}_{\bar{u}}c_{u}\bar{c}_{\bar{v}}c_{v}\rangle=-2\frac{\partial\Omega}{\partial U_{u\bar{u}v\bar{v}}}\Bigg|_{J}\label{eq:G2_def}
\end{align}

$G_{2}^{\mathrm{nc}}$ contains disconnected as well as connected
terms (hence the superscript nc for ``non connected''). We further
define the connected four-point correlator as:
\begin{align}
G_{2,\bar{u}u\bar{v}v} & \equiv G_{2,\bar{u}u\bar{v}v}^{\mathrm{nc}}+G_{u\bar{u}}G_{v\bar{v}}-G_{v\bar{u}}G_{u\bar{v}}\label{eq:G2_conn_def}
\end{align}

\subsection{Two-particle irreducible formalism}

\subsubsection{Legendre transformation}

By performing a Legendre transformation of the free energy with respect
to the bilinear sources $J$, one gets the Baym-Kadanoff\cite{Baym1961,Baym1962}
functional:
\begin{equation}
\Gamma_{2}[G,U]\equiv\Omega[J,U]+\mathrm{Tr}JG\label{eq:Gamma2_def}
\end{equation}

$\Gamma_{2}$ falls into two parts:
\begin{equation}
\Gamma_{2}[G,U]=\Gamma_{2,0}[G]+\Phi_{\mathrm{LW}}[G,U]\label{eq:Gamma2_decomp}
\end{equation}

$\Phi_{\mathrm{LW}}$ is the Luttinger-Ward functional\cite{Luttinger1960}:
it is made up of all two particle-irreducible (2PI) diagrams, namely
all diagrams which do not fall apart if any two of their lines are
cut open. The non-interacting contribution, $\Gamma_{2,0}$, is given
by
\begin{equation}
\Gamma_{2,0}[G]=-\mathrm{Tr}\log\left[G^{-1}\right]+\mathrm{Tr}\left[\left(G^{-1}-G_{0}^{-1}\right)G\right]\label{eq:baym_kadanoff}
\end{equation}

The physical solution is obtained by setting the source term $J$
to zero, i.e. by requiring the stationarity of $\Gamma_{2}$ stemming
from the reciprocity relation: 
\begin{equation}
\frac{\partial\Gamma_{2}}{\partial G}=J=0\label{eq:stationarity_Gamma2}
\end{equation}
This condition is equivalent (through Eqs (\ref{eq:Gamma2_decomp}-\ref{eq:baym_kadanoff}))
to the Dyson equation
\begin{equation}
\Sigma_{\bar{u}v}=G_{0,\bar{u}v}^{-1}-G_{\bar{u}v}^{-1}\label{eq:Dyson}
\end{equation}

where the self-energy $\Sigma$ is defined as the derivative of $\Phi_{\mathrm{LW}}$
with respect to $G$:

\begin{equation}
\Sigma_{\bar{u}v}=\frac{\partial\Phi_{\mathrm{LW}}}{\partial G_{v\bar{u}}}\Bigg|_{U}\label{eq:Sigma_LW}
\end{equation}

The 2PI functional allows to generate self-consistent approximation
methods by restricting $\Phi_{\mathrm{LW}}[G,U]$ to a (computable)
class of diagrams. Choosing a particular approximate form of $\Phi_{\mathrm{LW}}$
determines an approximate form of $\Sigma[G,U]$ and hence $G$ via
Dyson's equation (although there are some caveats to this procedure,
as recently demonstrated\cite{Kozik}).

\subsubsection{DMFT as an expansion of $\Phi_{\mathrm{LW}}$ around the atomic limit\label{subsec:The-DMFT-as-local_exp}}

Let us first briefly review the DMFT construction. DMFT consists in
approximating $\Phi_{\mathrm{LW}}$ by an expansion around the atomic
limit:\cite{Georges1996}
\begin{align}
 & \Phi_{\mathrm{LW}}^{\mathrm{DMFT}}[G_{\mathbf{R}\mathbf{R}'},U_{\mathbf{RR''R'''R''''}}]\nonumber \\
 & \equiv\sum_{\mathbf{R}}\Phi_{\mathrm{LW}}[G_{\mathbf{R}\mathbf{R}},U_{\mathbf{RRRR}}]\label{eq:DMFT_approx_phi}
\end{align}

In the right-hand side, $\Phi_{\mathrm{LW}}[G_{\mathbf{R}\mathbf{R}}]$
is shorthand for $\Phi_{\mathrm{LW}}[G_{\mathbf{R}\mathbf{R}}\delta_{\mathbf{R}\mathbf{R}'}]$
(and similarly for $U$). The form of this approximation shows that
DMFT is best suited for local interactions ($U_{\mathbf{RR''R'''R''''}}=U\delta_{\mathbf{RR''R'''R''''}}$).\footnote{In the DMFT approximation, nonlocal interactions only contribute at
the Hartree level.\cite{Muller-Hartmann1989}}

As a result, the DMFT self-energy is local:
\begin{equation}
\Sigma_{\mathbf{RR'}}^{\mathrm{DMFT}}(i\omega)=\Sigma_{\mathbf{RR}}(i\omega)\delta_{\mathbf{R}\mathbf{R}'}\label{eq:Sigma_DMFT_local}
\end{equation}

Here, $i\omega$ denotes a fermionic Matsubara frequency.

The resummation of the infinite class of local diagrams in (\ref{eq:DMFT_approx_phi})
is done by the following construction.

First, one introduces the following auxiliary impurity model:

\begin{align}
S_{\mathrm{imp}}^{\mathrm{DMFT}} & =-\iint_{\tau\tau'}\sum_{\sigma\sigma'}\bar{c}_{\tau\sigma}\left[\mathcal{G}^{-1}(\tau-\tau')\right]_{\sigma\sigma'}c_{\tau'\sigma'}\;\label{eq:S_imp_DMFT}\\
 & \;\;+\frac{1}{2}\int_{\tau}\sum_{\substack{\sigma_{1}\sigma_{2}\\
\sigma_{3}\sigma_{4}
}
}U_{\sigma_{1}\sigma_{2}\sigma_{3}\sigma_{4}}\bar{c}_{\tau\sigma_{1}}c_{\tau\sigma_{2}}\bar{c}_{\tau\sigma_{3}}c_{\tau\sigma_{4}}\nonumber 
\end{align}

Its Luttinger-Ward functional $\Phi_{\mathrm{LW}}^{\mathrm{imp}}$
is the same as the summand in the right-hand side of Eq. (\ref{eq:DMFT_approx_phi}).
Note that $\Phi_{\mathrm{LW}}^{\mathrm{imp}}$ depends on the full
propagator $G$ and bare interaction $U$, not on the non-interacting
propagator $\mathcal{G}$. 

Second, one adjusts the non-interacting propagator $\mathcal{G}$
of the auxiliary model such that
\begin{equation}
G_{\mathrm{imp}}[\mathcal{G}](i\omega)=G_{\mathbf{R}\mathbf{R}}(i\omega)\label{eq:sc_cond_DMFT}
\end{equation}

where the notation $[\mathcal{G}${]} means that $G_{\mathrm{imp}}$
depends on $\mathcal{G}$ through the solution of the impurity model,
Eq.(\ref{eq:S_imp_DMFT}). $\mathcal{G}$ can be regarded as a Lagrange
multiplier to enforce the constraint (\ref{eq:sc_cond_DMFT}).\cite{Georges2008} 

Finally, if Eq.(\ref{eq:sc_cond_DMFT}) is satisfied, then
\[
\Phi_{\mathrm{LW}}^{\mathrm{imp}}[G_{\mathrm{imp}},U]=\Phi_{\mathrm{LW}}[G_{\mathbf{RR}},U]
\]
and therefore Eq. (\ref{eq:Sigma_DMFT_local}) implies that 
\begin{equation}
\Sigma^{\mathrm{DMFT}}(\mathbf{k},i\omega)=\Sigma_{\mathrm{imp}}(i\omega)\label{eq:Sigma_approx_DMFT}
\end{equation}

The determination of the $\mathcal{G}$ fulfilling (\ref{eq:sc_cond_DMFT})
is usually done in an iterative fashion. We emphasize that in this
construction, $U$ is the same in the lattice model and in the impurity
model. Cluster DMFT methods,\cite{Lichtenstein2000,Kotliar2001,Hettler1998,Hettler1999,Maier2005a}
which consist in introducing an extended (i.e multi-site) impurity
model instead of Eq. (\ref{eq:S_imp_DMFT}), provide a systematic
expansion beyond DMFT.

\subsection{A reminder on vertex functions and the parquet formalism\label{subsec:Reminder:-definition-of}}

In this section, we give a reminder of the parquet equations\cite{Landau1954,Dominicis1964a}
so as to fix our notations (which are similar to those used in Refs
\onlinecite{Yang2009, Rohringer2012, Schafer2013, Rohringer2013a, Tam2013}). 

\begin{figure}

\begin{centering}
\includegraphics[width=0.8\columnwidth]{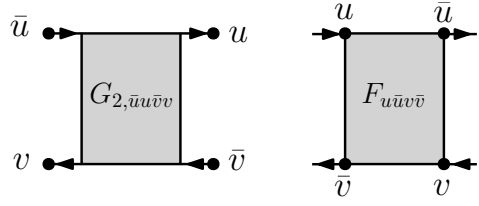}
\par\end{centering}
\caption{Graphical representation of the 4-point functions\label{fig:Graphical-representation-of}}

\end{figure}

The fully reducible vertex $F$ is defined as the amputated, connected
four-point function: 
\begin{equation}
F_{u\bar{u}v\bar{v}}\equiv G_{\bar{a}u}^{-1}G_{\bar{u}a}^{-1}G_{2,\bar{a}a\bar{b}b}G_{\bar{b}v}^{-1}G_{\bar{v}b}^{-1}\label{eq:F_def}
\end{equation}

$F$ contains all connected diagrams with two outgoing and two ingoing
entries. We note that $G_{2}$ and $F$ are of slightly different
nature: $F$ is of the ``vertex'' type (it is amputated, i.e. its
external points correspond to bare vertices), while $G_{2}$ is a
``correlator'' (it is not amputated, i.e. its external points correspond
to propagator ends). In diagrams, ``vertices'' can only be connected
to ``correlators'', and reciprocally. $G_{2}$ and $F$ are shown
graphically in Fig. \ref{fig:Graphical-representation-of}. 

We next define the irreducible vertex in channel $r$, $\Gamma^{r}$,
where $r=\mathrm{ph},\overline{\mathrm{ph}},\mathrm{pp}$. The irreducible
vertex in the particle-hole channel, $\Gamma^{\mathrm{ph}}$ (resp.
irreducible vertex in the horizontal particle-hole channel, $\Gamma^{\mathrm{\overline{ph}}}$),
contains all diagrams that do not fall apart if two horizontal (resp.
vertical) counterpropagating propagators are cut open. Similarly,
the irreducible vertex in the particle-particle channel, $\Gamma^{\mathrm{pp}}$,
contains all diagrams which do not fall apart when two propagators
going in the same direction are cut open. 

\begin{figure}
\begin{centering}
\includegraphics[width=1\columnwidth]{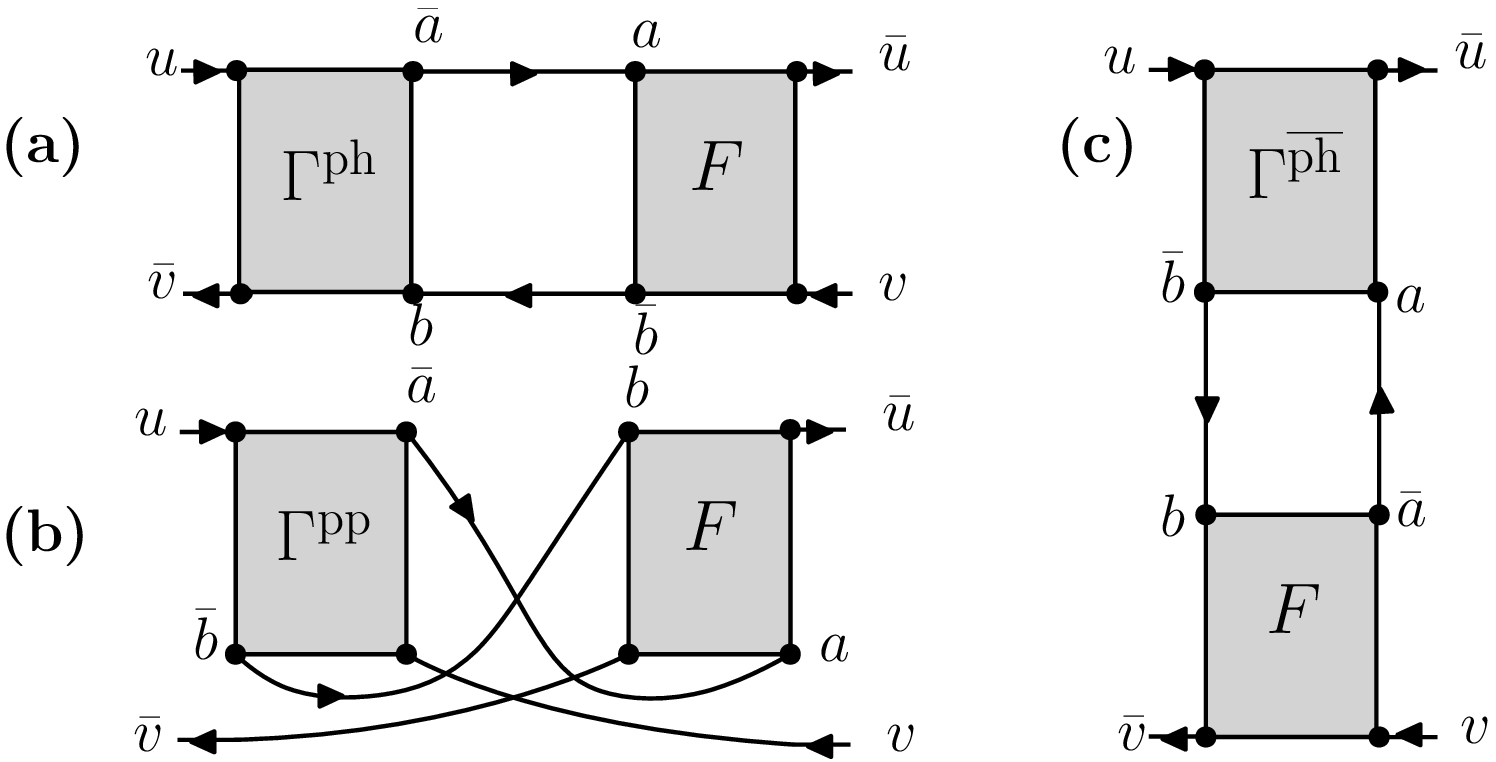}
\par\end{centering}
\caption{Graphical representation of the reducible vertex $\Phi^{r}$ in the
three channels. (a) $\Phi^{\mathrm{ph}}$ (b) $\Phi^{\mathrm{pp}}$
(c) $\Phi^{\overline{\mathrm{ph}}}$.\label{fig:Graphical-representation-of-Phi}}
\end{figure}

These diagrammatic definitions imply that $F$ and $\Gamma^{r}$ are
related by the Bethe-Salpether equation:
\begin{eqnarray}
F_{u\bar{u}v\bar{v}} & = & \Gamma_{u\bar{u}v\bar{v}}^{r}+\Phi_{u\bar{u}v\bar{v}}^{r}\label{eq:BSE}
\end{eqnarray}

where\begin{subequations}
\begin{align}
\Phi_{u\bar{u}v\bar{v}}^{\mathrm{ph}} & =\Gamma_{u\bar{a}b\bar{v}}^{\mathrm{ph}}G_{a\bar{a}}G_{b\bar{b}}F_{a\bar{u}v\bar{b}}\label{eq:Phi_ph}\\
\Phi_{u\bar{u}v\bar{v}}^{\overline{\mathrm{ph}}} & =\Gamma_{u\bar{u}a\bar{b}}^{\mathrm{\overline{ph}}}G_{a\bar{a}}G_{b\bar{b}}F_{b\bar{a}v\bar{v}}\label{eq:Phi_ph_bar}\\
\Phi_{u\bar{u}v\bar{v}}^{\mathrm{pp}} & =\Gamma_{u\bar{a}v\bar{b}}^{\mathrm{pp}}G_{a\bar{a}}G_{b\bar{b}}F_{b\bar{u}a\bar{v}}\label{eq:Phi_pp}
\end{align}
\end{subequations}The function $\Phi^{r}$ is called the ``reducible
vertex in channel $r$''. These relations are illustrated in Fig.
\ref{fig:Graphical-representation-of-Phi}. 

Next, we define the (interacting and ``open'') bubble in channel
$r$, $\chi^{r}$, as:\begin{subequations}
\begin{align}
\chi_{\bar{u}u\bar{v}v}^{\mathrm{ph}} & \equiv G_{u\bar{u}}G_{v\bar{v}}\label{eq:chi_ph}\\
\chi_{\bar{u}u\bar{v}v}^{\mathrm{\overline{ph}}} & \equiv G_{v\bar{u}}G_{u\bar{v}}\label{eq:chi_ph_bar}\\
\chi_{\bar{u}u\bar{v}v}^{\mathrm{pp}} & \equiv G_{v\bar{u}}G_{u\bar{v}}\label{eq:chi_pp}
\end{align}
\end{subequations}

We now introduce a change of notation to unify the expressions (\ref{eq:Phi_ph}-\ref{eq:Phi_ph_bar}-\ref{eq:Phi_pp}).
For ``vertex'' functions $V_{u\bar{u}v\bar{v}}$ like $F$, $\Gamma^{r}$
and $\Phi^{r}$, we introduce the following three hatted functions
$\hat{V}_{r}$:\begin{subequations}
\begin{align}
\hat{V}_{\mathrm{ph},u\bar{v},\bar{u}v} & \equiv V_{u\bar{u}v\bar{v}}\label{eq:V_ph}\\
\hat{V}_{\overline{\mathrm{ph}},\bar{u}u,\bar{v}v} & \equiv V_{u\bar{u}v\bar{v}}\label{eq:V_ph_bar}\\
\hat{V}_{\mathrm{pp},uv,\bar{u}\bar{v}} & \equiv V_{u\bar{u}v\bar{v}}\label{eq:V_pp}
\end{align}

\end{subequations}Here, the \emph{subscript} $r$ defines the pairing
of the four indices. This is to be distinguished from \emph{superscripts}
(like in $\Phi^{r}$), which denote an intrinsic dependence on the
channel. Thus $\hat{F}_{r}$, which is ``$F$ in the $r$ notation'',
depends on $r$ (whereas $F$ does not intrinsically depend on $r$).
$\hat{\Phi}_{r'}^{r}$ is ``$\Phi^{r}$ in the $r'$ notation'',
it depends on $r'$ (subscript) through the notation and intrinsically
on $r$ (superscript). For ``correlator'' functions $C_{\bar{u}u\bar{v}v}$
(like $G_{2}$ and $\chi^{r}$), likewise, we introduce the following
three hatted functions: \begin{subequations}

\begin{align}
\hat{C}_{\mathrm{ph},\bar{u}v,u\bar{v}} & \equiv C_{\bar{u}u\bar{v}v}\label{eq:C_ph}\\
\hat{C}_{\overline{\mathrm{ph}},\bar{v}v,\bar{u}u} & \equiv C_{\bar{u}u\bar{v}v}\label{eq:C_ph_bar}\\
\hat{C}_{\mathrm{pp},\bar{u}\bar{v},uv} & \equiv C_{\bar{u}u\bar{v}v}\label{eq:C_pp}
\end{align}

\end{subequations}

With these notations, Eqs (\ref{eq:Phi_ph}-\ref{eq:Phi_ph_bar}-\ref{eq:Phi_pp})
become a simple matrix product:

\begin{equation}
\hat{\Phi}_{r,\alpha\beta}^{r}\equiv\hat{\Gamma}_{r,\alpha\gamma}^{r}\hat{\chi}_{r,\gamma\delta}^{r}\hat{F}_{r\delta\beta}\label{eq:Phi_def}
\end{equation}

Here, Greek indices denote the channel-dependent combination of two
fermionic indices. They only make sense with a subscript $r$ to specify
which pairing of indices is chosen. 

We also note (see Appendix \ref{sec:Relation-between-} for a proof)
for further reference that we have, for all $r$:
\begin{equation}
\hat{G}_{2}=\hat{\chi}^{r}\hat{F}\hat{\chi}^{r}\label{eq:G2_XFX}
\end{equation}

The passage from the notation in channel $r$ to the notation in channel
$r'$ is performed via a tensor $\zeta_{\alpha\beta,\gamma\delta}^{r'r}$
defined by the following transformation of ``correlators'':
\begin{equation}
\hat{C}_{r',\alpha\beta}=\zeta_{\alpha\beta,\gamma\delta}^{r'r}\hat{C}_{r,\gamma\delta}\label{eq:zeta_def}
\end{equation}

Here, we do not sum over $r$ and $r'$. Some basic properties of
this tensor are summarized in Appendix \ref{sec:Properties-of-the-zeta-tensor}.
We further note that the trace of two operators which do not intrinsically
depend on $r$ does not depend on the choice of notation, i.e. 
\begin{equation}
\mathrm{Tr}\hat{C}\hat{V}=\hat{C}_{r,\alpha\beta}\hat{V}_{r,\beta\alpha}=\hat{C}_{r',\gamma\delta}\hat{V}_{r',\delta\gamma}\label{eq:trace_indep_r}
\end{equation}

The transformation from $r$ notation to $r'$ notation for vertex
functions follows from this property:\footnote{Indeed, using Eq. (\ref{eq:closure_zeta}) of Appendix \ref{sec:Properties-of-the-zeta-tensor},
one can check:
\[
\hat{C}_{r,\alpha\beta}\hat{V}_{r,\beta\alpha}=\zeta_{\alpha\beta,\gamma\delta}^{rr'}\hat{C}_{r',\gamma\delta}\zeta_{\bar{\delta}\bar{\gamma},\alpha\beta}^{r'r}\hat{V}_{r',\bar{\gamma}\bar{\delta}}=\hat{C}_{r',\gamma\delta}\hat{V}_{r',\delta\gamma}
\]
}

\begin{equation}
\hat{V}_{r',\alpha\beta}=\zeta_{\delta\gamma,\beta\alpha}^{rr'}\hat{V}_{r,\gamma\delta}\label{eq:zeta_vertex}
\end{equation}

In the above expressions, Einstein summation is performed only on
the Greek indices. For the same reason as above, the inverse of correlators
transform like vertex functions.

The Bethe-Salpether equation Eq. (\ref{eq:BSE}) can now be formally
inverted. For all $r$'s, we have:
\begin{equation}
\hat{\Gamma}_{r}^{r}=\hat{F}_{r}(\hat{\mathbf{1}}+\hat{\chi}_{r}^{r}\hat{F}_{r})^{-1}\label{eq:inverse_BSE}
\end{equation}

where inversion is performed in the space of Greek indices.

Finally, we define the fully irreducible vertex $\Lambda$. It contains
all diagrams that are irreducible in the ph, $\overline{\mathrm{ph}}$
and pp channels. It thus obeys the relation:
\begin{equation}
F=\Lambda+\sum_{r}\Phi^{r}\label{eq:F_decomp}
\end{equation}

Combining (\ref{eq:BSE}) and (\ref{eq:F_decomp}) yields:
\begin{equation}
\Gamma^{r}=\Lambda+\sum_{r'\neq r}\Phi^{r'}\label{eq:Gamma_r_decomp}
\end{equation}

The parquet equations are obtained by using the definition of $\Phi^{r}$,
Eq. (\ref{eq:Phi_def}), and replacing $\Gamma^{r}$ and $F$ using
(\ref{eq:F_decomp}) and (\ref{eq:Gamma_r_decomp}):
\begin{equation}
\hat{\Phi}_{r}^{r}=\left(\hat{\Lambda}_{r}+\sum_{r'\neq r}\hat{\Phi}_{r}^{r'}\right)\hat{\chi}_{r}^{r}\left(\hat{\Lambda}_{r}+\sum_{r'}\hat{\Phi}_{r}^{r'}\right)\label{eq:parquet}
\end{equation}

The parquet equations relate $\Lambda$ and $\Phi^{r}$ (at fixed
$\chi^{r}$ i.e. fixed $G$), and thus {[}through Eqs(\ref{eq:F_decomp}-\ref{eq:G2_XFX}){]}
$\Lambda$ to $G_{2}$. They couple the three channels (the passage
from $\hat{\Phi}_{r}^{r}$ to $\hat{\Phi}_{r'}^{r}$ is given by Eq.(\ref{eq:zeta_def})).
Conversely, the inverse parquet equations consist in computing $\Gamma^{r}$
and $\Phi^{r}$ from a given $G_{2}$ or $F$ {[}via Eq. (\ref{eq:inverse_BSE})
and (\ref{eq:BSE}){]}, and eventually {[}through (\ref{eq:F_decomp}){]}
$\Lambda$. They do not couple the three channels and are as such
much easier to solve than the direct parquet equations.

The first contribution to $\Lambda$ is the bare interaction $U$.
It is thus natural to define the correction of $\Lambda$ beyond $U$
as:

\begin{equation}
\delta\Lambda\equiv\Lambda-U\label{eq:Lambda_Lambda_tilde}
\end{equation}

The lowest-order diagram of $\delta\Lambda$ is of order $U^{4}$.
It is shown in Fig. \ref{fig:Left:-simplest-diagramKappa4}, right
panel. 

One can now observe that the parquet equations formally relate the
bare interactions $U$, the nontrivial contribution to the fully irreducible
vertex, $\delta\Lambda$, and the (fully reducible) two-particle correlator
$G_{2}$ (the functions $\Gamma^{r}$ and $\Phi^{r}$ can be regarded
as by-standers). In that sense, they are analogous to the Dyson equations,
which relate the bare correlator $G_{0}$, the irreducible contribution
or self-energy $\Sigma$ and the (full) one-particle correlator $G$.

We note that in a single-orbital context, all the above-mentioned
four-point functions depend on three momenta and three frequencies
in the time- and space-translation invariant case, as well as orbital
and spin indices, e.g.
\[
\Lambda_{\sigma_{1}\sigma_{2}\sigma_{3}\sigma_{4}}(\mathbf{k},\mathbf{k}',\mathbf{q},i\omega,i\omega',i\Omega)
\]

Further simplifications of the spin structure arise in SU(2) invariant
problems (see e.g. Ref. \onlinecite{Rohringer2013a} for more details).

\subsection{Four-particle irreducible formalism}

Here, we introduce (subsection \ref{subsec:Legendre-transformation})
the Legendre transform of $\Gamma_{2}[G,U]$ with respect to the quartic
sources, as well as its irreducible part $\mathcal{K}_{4}$ and its
properties. We then show that the approximation $\mathcal{K}_{4}=0$
corresponds to the parquet approximation (subsection \ref{subsec:Self-consistent-approx-Kappa4}),
and finally prove that approximations on $\mathcal{K}_{4}$ preserve
the consistency of the self-energy given as the derivative as the
Luttinger-Ward functional or by the Schwinger-Dyson equation (subsection
\ref{subsec:Consistency-of-the-selfenergy}). 

\subsubsection{Legendre transformation\label{subsec:Legendre-transformation}}

We define the Legendre transform of $\Gamma_{2}$ with respect to
$U$:
\begin{equation}
\Gamma_{4}[G,G_{2}]\equiv\Gamma_{2}[G,U]+\frac{1}{2}\hat{U}_{r,\alpha\beta}\hat{G}_{2,r,\beta\alpha}^{\mathrm{nc}}\label{eq:Gamma4_def}
\end{equation}

$\Gamma_{4}$ is a functional of $G$ and $G_{2}$ (or equivalently
of $G$ and $F$ via (\ref{eq:G2_XFX})) because $U$ is a functional
of $G$ and $G_{2}$ through the relation: 
\begin{equation}
G_{2,\bar{u}u\bar{v}v}^{\mathrm{nc}}[G,U]=-2\frac{\partial\Gamma_{2}}{\partial U_{u\bar{u}v\bar{v}}}\Bigg|_{G}\label{eq:G2_nc_der_Gamma2}
\end{equation}

which follows from Eq. (\ref{eq:G2_def}) and the properties of the
Legendre transform. Note that the second term in the definition of
$\Gamma_{4}$ does not depend on $r$ due to Eq. (\ref{eq:trace_indep_r}).
The passage from a functional of the bare interaction $U$ to a functional
of the $G_{2}$ (or $F$) has first been proposed in Ref. \onlinecite{Dominicis1964a},
and is also investigated in Ref. \onlinecite{VanLeeuwen2006}.

$\Gamma_{4}$ is the entropy of the system, up to a minus sign and
a shift of the source $J-t\leftarrow J$ (where $t_{\bar{u}v}$ is
the hopping integral in the quadratic part of the Hamiltonian $H$
corresponding to the action defined in Eq.(\ref{eq:ee_action})).
Indeed, Eqs (\ref{eq:Gamma2_def}-\ref{eq:Gamma4_def}) give the relation:
\begin{equation}
T\Omega=\langle H\rangle-T(-\Gamma_{4})\label{eq:entropy_and_Gamma4}
\end{equation}

where $T$ is the temperature.

Finally, from Eq. (\ref{eq:Gamma4_def}) and (\ref{eq:G2_def}) follows
the reciprocity relation:
\begin{equation}
\frac{1}{2}\hat{U}_{r,\alpha\beta}=\frac{\partial\Gamma_{4}}{\partial\hat{G}_{2,r,\beta\alpha}}\Bigg|_{G}\label{eq:reciprocity_Gamma4}
\end{equation}

Following Ref \onlinecite{Dominicis1964a}, we define the following
functional:\footnote{Up to notations and factors, this corresponds to Eq. (60) of Ref.
\onlinecite{Dominicis1964a} and the functional $L'[G,F]$ of Ref.
\onlinecite{VanLeeuwen2006}}
\begin{eqnarray}
 &  & \mathcal{K}_{4}[G,G_{2}]\equiv\Phi_{\mathrm{LW}}[G,U]+\frac{1}{2}\hat{U}_{r,\alpha\beta}\hat{G}_{2,r,\beta\alpha}^{\mathrm{nc}}\nonumber \\
 &  & \;\;\;\;\;+\frac{1}{2}\hat{F}_{r,\alpha\beta}\hat{G}_{2,r,\beta\alpha}-\frac{1}{2}\sum_{r}\Theta^{r}[G,G_{2}]\label{eq:Kappa4_def}
\end{eqnarray}

with
\begin{align}
 & \Theta^{r}[G,G_{2}]\equiv\label{eq:def_Theta}\\
 & -\mathrm{Tr}\left[\ln\left(\hat{\mathbf{1}}+\hat{G}_{2,r}\left(\hat{\chi}_{r}^{r}\right)^{-1}\right)-\hat{G}_{2,r}\left(\hat{\chi}_{r}^{r}\right)^{-1}\right]\nonumber 
\end{align}

Thus, using Eqs (\ref{eq:Gamma2_decomp}) and (\ref{eq:Kappa4_def})
in Eq (\ref{eq:Gamma4_def}), $\Gamma_{4}$ can be written as:
\begin{eqnarray}
 &  & \Gamma_{4}[G,G_{2}]=\Gamma_{2,0}[G]-\frac{1}{2}\hat{F}_{r,\alpha\beta}\hat{G}_{2,r,\beta\alpha}\nonumber \\
 &  & \;+\frac{1}{2}\sum_{r}\Theta^{r}[G,G_{2}]+\mathcal{K}_{4}[G,G_{2}]\label{eq:Gamma4_decomp}
\end{eqnarray}
which highlights the dependence of $\Gamma_{4}$ on $G$ and $G_{2}$,
and its decomposition into explicit terms (the first three terms)
and a nontrivial term, $\mathcal{K}_{4}$.

\begin{figure}
\begin{centering}
\includegraphics[width=0.9\columnwidth]{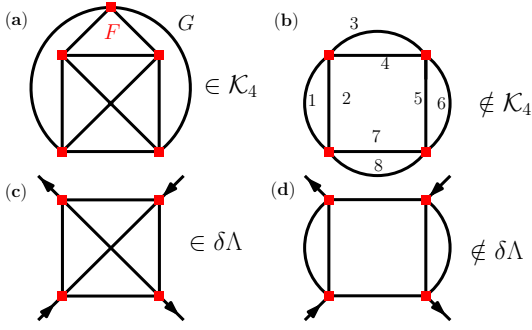}
\par\end{centering}
\caption{(a) simplest diagram of $\mathcal{K}_{4}$ (b) an example of 4-particle-reducible
diagram (c) simplest diagram of $\delta\Lambda$ (d) an example of
reducible 4-leg diagram. Lines denote $G$, while red squares denote
$F$. \label{fig:Left:-simplest-diagramKappa4}}
\end{figure}

This definition of $\mathcal{K}_{4}$ ensures that $\mathcal{K}_{4}$
can be represented diagrammatically by the set of all four-particle-irreducible
(4PI) diagrams, as shown in Ref. \onlinecite{Dominicis1964a} in the
case of bosonic fields. A diagram is said to be 4PI if for any set
of four lines whose cutting leads to the separation of the diagram
into two disconnected pieces, one and only one of the pieces is a
simple four-leg vertex $F$. The lowest-order diagram of $\mathcal{K}_{4}$
is shown in Fig. \ref{fig:Left:-simplest-diagramKappa4}(a). In Fig.
\ref{fig:Left:-simplest-diagramKappa4} (b), for instance, cutting
lines 3,4, 7 and 8 leads to the separation of the diagram into two
disconnected pieces, none of which is a simple four-leg vertex $F$;
therefore, this diagram is not 4PI.

Moreover, with definition (\ref{eq:Kappa4_def}), we are to show (in
the next subsection) that the fully irreducible vertex $\Lambda$
(or $\delta\Lambda$) derives from $\mathcal{K}_{4}$:

\begin{equation}
\delta\hat{\Lambda}_{r,\alpha\beta}=-2\frac{\partial\mathcal{K}_{4}[G,G_{2}]}{\partial\hat{G}_{2,r,\beta\alpha}}\Bigg|_{G}\label{eq:Lambda_tilde_def}
\end{equation}

This is illustrated in Figure \ref{fig:Left:-simplest-diagramKappa4}(c),
which corresponds to the graphical derivative of the diagram of panel
(a) of \ref{fig:Left:-simplest-diagramKappa4}. This property can
be used to devise various approximations at the 4PI level, as will
be illustrated in sections \ref{subsec:Self-consistent-approx-Kappa4}
and \ref{subsec:Quadrilex-approximation}.

Eq. (\ref{eq:Lambda_tilde_def}) remarkably parallels Eq. (\ref{eq:Sigma_LW})
of the previous section. At the 2PI level, the stationarity of $\Gamma_{2}$
{[}Eq (\ref{eq:stationarity_Gamma2}){]} is equivalent to the fulfillment
of Dyson's equation {[}Eq(\ref{eq:Dyson}){]} between $G_{0}$, $G$,
and $\Sigma$, the derivative of the 2PI functional, $\Phi_{\mathrm{LW}}$.
Similarly, at the 4PI level, the stationarity of $\Gamma_{4}$ {[}Eq.
(\ref{eq:reciprocity_Gamma4}){]} is equivalent to the fulfillment
of the parquet equations between $U$, $G_{2}$ and $\delta\Lambda$,
the derivative of the 4PI functional $\mathcal{K}_{4}$. 

\subsubsection{Proof of Eq. (\ref{eq:Lambda_tilde_def})}

Starting from Eq.(\ref{eq:reciprocity_Gamma4}), we use Eqs (\ref{eq:Gamma4_def})-(\ref{eq:Gamma4_decomp})
as well as the property (deduced from Eq.(\ref{eq:zeta_def})):

\begin{equation}
\frac{\partial\hat{G}_{2,r',\kappa\eta}}{\partial\hat{G}_{2,r,\beta\alpha}}=\zeta_{\kappa\eta,\beta\alpha}^{r'r}\label{eq:derivative_zeta}
\end{equation}

and we remember that $F$ is just the amputated $G_{2}$ (Eq.(\ref{eq:F_def}),
i.e the ``$FG_{2}$'' term contains two $G_{2}$'s at fixed $G$),
to write: 
\begin{eqnarray}
 &  & \frac{1}{2}\hat{U}_{r,\alpha\beta}=\frac{\partial\mathcal{K}_{4}}{\partial\hat{G}_{2,r,\beta\alpha}}\Big|_{G}-\hat{F}_{r,\alpha\beta}\nonumber \\
 &  & +\frac{1}{2}\sum_{r'}\zeta_{\kappa\eta,\beta\alpha}^{r'r}\frac{\partial\Theta^{r'}[G,G_{2}]}{\partial\hat{G}_{2,r',\kappa\eta}}\label{eq:interm_U}
\end{eqnarray}

Using the chain rule and Eqs. (\ref{eq:G2_XFX}-\ref{eq:inverse_BSE}),
the last term evaluates to:

\begin{eqnarray}
 &  & \frac{\partial\Theta^{r}[G,G_{2}]}{\partial\hat{G}_{2,r,\kappa\eta}}\nonumber \\
 &  & =-\left[\left(\hat{\chi}_{r}^{r}\right)^{-1}\left(\hat{\mathbf{1}}+\hat{G}_{2}\left(\hat{\chi}_{r}^{r}\right)^{-1}\right)^{-1}-\left(\hat{\chi}_{r}^{r}\right)^{-1}\right]_{\eta\kappa}\nonumber \\
 &  & =-\left[\left(\hat{\chi}_{r}^{r}\right)^{-1}-\left(\hat{\chi}_{r}^{r}\right)^{-1}\left(\hat{\mathbf{1}}+\hat{G}_{2,r}\left(\hat{\chi}_{r}^{r}\right)^{-1}\right)\right]_{\eta\alpha}\nonumber \\
 &  & \;\;\;\;\times\left[\hat{\mathbf{1}}+\hat{G}_{2,r}\left(\hat{\chi}_{r}^{r}\right)^{-1}\right]_{\alpha\kappa}^{-1}\nonumber \\
 &  & =-\left[-\left(\hat{\chi}_{r}^{r}\right)^{-1}\hat{G}_{2,r}\left(\hat{\chi}_{r}^{r}\right)^{-1}\right]_{\eta\alpha}\left[\hat{\mathbf{1}}+\hat{G}_{2,r}\left(\hat{\chi}_{r}^{r}\right)^{-1}\right]_{\alpha\kappa}^{-1}\nonumber \\
 &  & =\hat{F}_{r,\eta\alpha}\left(\hat{\mathbf{1}}+\hat{\chi}_{r}^{r}\hat{F}_{r}\right)_{\alpha\kappa}^{-1}\nonumber \\
 &  & =\hat{\Gamma}_{r,\eta\kappa}^{r}\label{eq:derivative_of_the_log}
\end{eqnarray}

Hence, using Eqs. (\ref{eq:BSE}-\ref{eq:F_decomp}), we find, using
Eq. (\ref{eq:zeta_vertex}) and multiplying (\ref{eq:interm_U}) by
2:\begin{subequations}

\begin{eqnarray}
\hat{U}_{r,\alpha\beta} & = & 2\frac{\partial\mathcal{K}_{4}}{\partial\hat{G}_{2,r,\beta\alpha}}-2\hat{F}_{r,\alpha\beta}+\sum_{r'}\zeta_{\kappa\eta,\beta\alpha}^{r'r}\hat{\Gamma}_{r',\eta\kappa}^{r'}\nonumber \\
 & = & 2\frac{\partial\mathcal{K}_{4}}{\partial\hat{G}_{2,r,\beta\alpha}}-2\hat{F}_{r,\alpha\beta}+\sum_{r'}\hat{\Gamma}_{r,\alpha\beta}^{r'}\label{eq:U_interm_0}\\
 & = & 2\frac{\partial\mathcal{K}_{4}}{\partial\hat{G}_{2,r,\beta\alpha}}-2\hat{F}_{r,\alpha\beta}+\sum_{r'}\left(\hat{F}_{r,\alpha\beta}-\hat{\Phi}_{r,\alpha\beta}^{r'}\right)\nonumber \\
 & = & 2\frac{\partial\mathcal{K}_{4}}{\partial\hat{G}_{2,r,\beta\alpha}}+\hat{F}_{r,\alpha\beta}-\sum_{r'}\hat{\Phi}_{r,\alpha\beta}^{r'}\nonumber \\
 & = & 2\frac{\partial\mathcal{K}_{4}}{\partial\hat{G}_{2,r,\beta\alpha}}+\hat{\Lambda}_{r,\alpha\beta}\label{eq:U_interm}
\end{eqnarray}

\end{subequations}In the last step, we have used Eq. (\ref{eq:F_decomp}).
By identification with Eq. (\ref{eq:Lambda_Lambda_tilde}), we find
the final result, Eq. (\ref{eq:Lambda_tilde_def}).

\subsubsection{The parquet approximation: $\mathcal{K}_{4}=0$\label{subsec:Self-consistent-approx-Kappa4}}

The most trivial approximation of $\mathcal{K}_{4}$, namely 
\begin{equation}
\mathcal{K}_{4}^{\mathrm{parquet\,app.}}=0\label{eq:parquet_approximation}
\end{equation}

corresponds to the parquet approximation. Indeed, Eq.(\ref{eq:parquet_approximation}),
combined with Eqs. (\ref{eq:Lambda_Lambda_tilde}-\ref{eq:Lambda_tilde_def}),
leads to
\begin{equation}
\Lambda^{\mathrm{parquet\;app.}}=U\label{eq:parquet_app_Lambda}
\end{equation}
By construction, this approximation is limited to the weak-coupling
regime since it neglects higher order terms. It has been recently
applied to the Hubbard model\cite{Yang2009}. We note that an alternative
functional view on the parquet approximation is proposed in Ref. \onlinecite{Janis1998}.

\begin{figure}

\begin{centering}
\includegraphics[width=1\columnwidth]{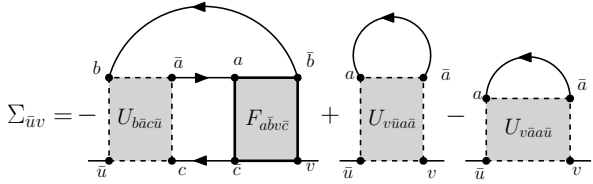}
\par\end{centering}
\caption{Schwinger-Dyson expression of the self-energy.\label{fig:Self-energy-SD}}

\end{figure}

\subsubsection{Consistency of the self-energy\label{subsec:Consistency-of-the-selfenergy}}

Any approximation of $\mathcal{K}_{4}[G,G_{2}]$ results in (i) an
approximate irreducible vertex $\delta\Lambda$ and, via the parquet
equations, approximate fully reducible vertex $F$ and (ii) an approximate
Luttinger-Ward functional {[}via. Eq. (\ref{eq:Kappa4_def}){]}. 

From here, there are \emph{a priori} two ways of computing the self-energy.
The first way is compute $\Sigma$ as the derivative of $\Phi_{\mathrm{LW}}$
with respect to $G$ {[}Eq. (\ref{eq:Sigma_LW}){]}. The second one
is to use the Schwinger-Dyson equation, an exact expression giving
$\Sigma$ as a function of $G$, $F$ and $U$ and illustrated in
Fig. \ref{fig:Self-energy-SD}:
\begin{equation}
\Sigma_{\bar{u}v}=-U_{b\bar{a}c\bar{u}}G_{a\bar{a}}G_{c\bar{c}}G_{b\bar{b}}F_{a\bar{b}v\bar{c}}+U_{v\bar{u}a\bar{a}}G_{a\bar{a}}-U_{v\bar{a}a\bar{u}}G_{a\bar{a}}\label{eq:Schwinger_Dyson}
\end{equation}

The last two terms correspond to the Hartree and Fock terms, respectively. 

We prove in Appendix \ref{sec:Self-Energy} that provided the approximation
on $\mathcal{K}_{4}[G,G_{2}]$ preserves its homogeneity properties,
both ways of computing the self-energy are equivalent. 

In the next subsection, we introduce, instead of the simple approximation
(\ref{eq:parquet_approximation}), an atomic approximation of $\mathcal{K}_{4}$.

\subsection{A quadruply irreducible local expansion: QUADRILEX\label{subsec:Quadrilex-approximation}}

\subsubsection{A local expansion of the 4PI functional\label{subsec:A-local-expansion}}

Similarly to DMFT, we propose to approximate the 4PI functional by
the atomic limit (and later by a cluster method):

\begin{align}
\mathcal{K}_{4}^{\mathrm{QUADRILEX}}[G_{\mathbf{R}_{1}\mathbf{R}_{2}},G_{2,\mathbf{R}_{1}\mathbf{R}_{2}\mathbf{R}_{3}\mathbf{R}_{4}}]\nonumber \\
\equiv\sum_{\mathbf{R}}\mathcal{K}_{4}[G_{\mathbf{RR}},G_{2,\mathbf{RRRR}}]\label{eq:quadrilex_approx}
\end{align}

To solve Eq. (\ref{eq:quadrilex_approx}), we propose to follow a
similar procedure as the one used in DMFT (see subsection \ref{subsec:The-DMFT-as-local_exp}),
by replacing $\Phi_{\mathrm{LW}}$ by $\mathcal{K}_{4}$:

First, we introduce the following model: 
\begin{align}
S_{\mathrm{imp}}^{\mathrm{QUADRILEX}} & =\label{eq:S_imp_quadrilex}\\
-\iint_{\tau\tau'} & \sum_{\sigma\sigma'}\bar{c}_{\tau\sigma}\left[\mathcal{G}^{-1}(\tau-\tau')\right]_{\sigma\sigma'}c_{\tau'\sigma'}\nonumber \\
+\frac{1}{2}\iiiint_{\substack{\tau_{1}\tau_{2}\\
\tau_{3}\tau_{4}
}
} & \sum_{\substack{\sigma_{1}\sigma_{2}\\
\sigma_{3}\sigma_{4}
}
}\mathcal{U}_{\sigma_{1}\sigma_{2}\sigma_{3}\sigma_{4}}^{\tau_{1},\tau_{2},\tau_{3},\tau_{4}}\bar{c}_{\tau_{1}\sigma_{1}}c_{\tau_{2}\sigma_{2}}\bar{c}_{\tau_{3}\sigma_{3}}c_{\tau_{4}\sigma_{4}}\nonumber 
\end{align}
This action describes an impurity embedded in a noninteracting bath
described by the field $\mathcal{G}$ and with dynamical interactions
$\mathcal{U}$ with three independent times. Its functional $\mathcal{K}_{4}[G,G_{2}]$
is the same as the summand of the right-hand side of Eq. (\ref{eq:quadrilex_approx}),
and does not depend on the non-interacting propagator $\mathcal{G}(i\omega)$
and bare interaction $\mathcal{U}(i\omega,i\omega',i\Omega)$. 

Second, we assume that one can adjust the non-interacting propagator
$\mathcal{G}$ and bare interaction $\mathcal{U}$ of the auxiliary
model such that\begin{subequations}
\begin{eqnarray}
G_{\mathrm{imp}}[\mathcal{G},\mathcal{U}](i\omega) & = & G_{\mathbf{RR}}(i\omega)\label{eq:G_sc}\\
G_{\mathrm{2,imp}}[\mathcal{G},\mathcal{U}](i\omega,i\omega',i\Omega) & = & G_{2,\mathbf{RRRR}}(i\omega,i\omega',i\Omega)\nonumber \\
\label{eq:G2_self_consistency}
\end{eqnarray}
\end{subequations}$\mathcal{G}$ and $\mathcal{U}$ can be thought
of as Lagrange multipliers to enforce the two above constraints. 

Finally, if we solve Eqs (\ref{eq:G_sc}-\ref{eq:G2_self_consistency}),
then
\[
\mathcal{K}_{4}^{\mathrm{imp}}[G_{\mathrm{imp}},G_{2,\mathrm{imp}}]=\mathcal{K}_{4}[G_{\mathbf{RR}},G_{2,\mathbf{RRRR}}]
\]
and therefore Eqs. (\ref{eq:Lambda_tilde_def})-(\ref{eq:quadrilex_approx})
imply that 
\begin{equation}
\delta\Lambda(\mathbf{k},\mathbf{k}',\mathbf{q},i\omega,i\omega',i\Omega)=\delta\Lambda_{\mathrm{imp}}(i\omega,i\omega',i\Omega)\label{eq:Lambda_tilde_approx}
\end{equation}

For simplicity, we will henceforth use the following shorthand notation
for 4-leg functions on the lattice:
\begin{equation}
X_{\mathrm{latt}}\equiv X_{\sigma_{1}\sigma_{2}\sigma_{3}\sigma_{4}}(\mathbf{k},\mathbf{k'},\mathbf{q},i\omega,i\omega',i\Omega)\label{eq:latt_notation}
\end{equation}

We point out that while DMFT is the approximation of $\Phi_{\mathrm{LW}}[G,U]$,
which depends on one full correlator, $G$, and has correspondingly
one ``dynamical bath'' $\mathcal{G}$, QUADRILEX, an approximation
of $\mathcal{K}_{4}[G,G_{2}]$, which depends on two full correlators,
$G$ and $G_{2}$, involves two dynamical mean fields $\mathcal{G}$
and $\mathcal{U}$ corresponding to two constraints, Eqs (\ref{eq:G_sc}-\ref{eq:G2_self_consistency}). 

\begin{figure*}

\begin{centering}
\includegraphics[width=0.75\textwidth]{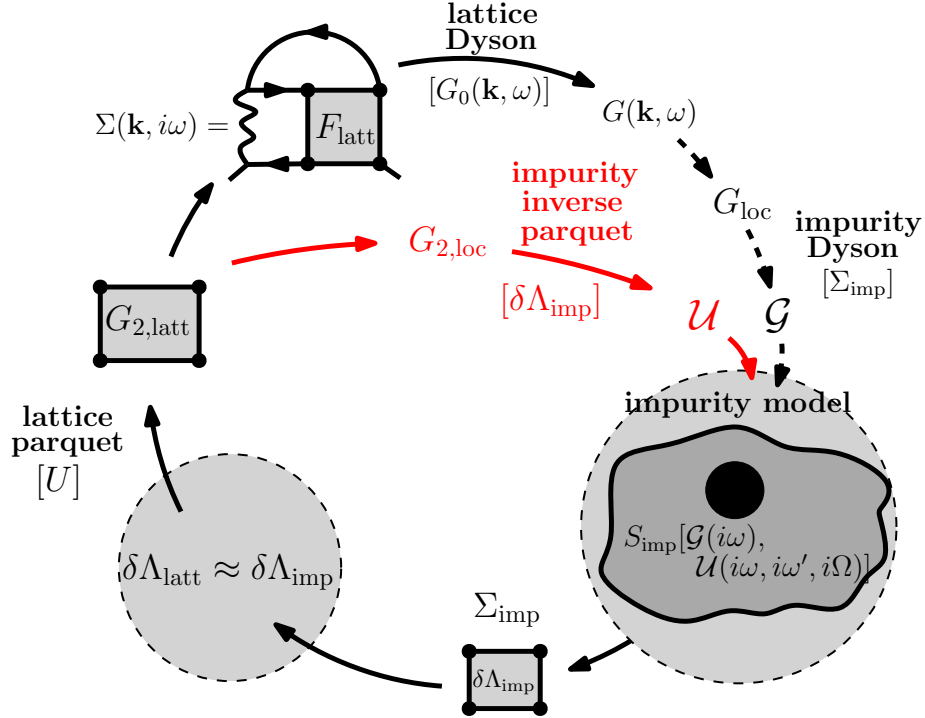}
\par\end{centering}
\caption{(color online) The QUADRILEX loop. The parts in black correspond to
the D$\Gamma$A in the parquet version (the dashed line shows how
D$\Gamma$A proposes to update the Weiss field $\mathcal{G}(i\omega)$).
The parts in red correspond to the additional steps necessary to update
the bare impurity vertex $\mathcal{U}(i\omega,i\omega',i\Omega)$.
For $\Sigma$ we show only the terms beyond Hartree-Fock. The terms
in square brackets denote the fixed terms in the solution of the Dyson/parquet
equations.\label{fig:The-quadrilex-loop}}

\end{figure*}

\subsubsection{QUADRILEX construction}

In dynamical mean field theory, the constraint Eq. (\ref{eq:G_sc})
is used together with the Dyson equation to determine the self-consistent
bath $\mathcal{G}$ from a given impurity self-energy $\Sigma_{\mathrm{imp}}$.
This is done as follows:

(i) the lattice self-energy $\Sigma(\mathbf{k},i\omega)$ is approximated
by the impurity self-energy $\Sigma_{\mathrm{imp}}$;

(ii) $\Sigma(\mathbf{k},i\omega)$ is plugged into the lattice Dyson
equation (at fixed $G_{0}(\mathbf{k},i\omega)$) to get $G(\mathbf{k},i\omega)$,
which is summed over the Brillouin zone to get $G_{\mathrm{loc}}(i\omega)$;

(iii) The impurity Dyson equation is inverted to get $\mathcal{G}$
from $\Sigma_{\mathrm{imp}}$ and $G_{\mathrm{loc}}$.

In QUADRILEX, the same procedure is applied to get the retarded interactions
$\mathcal{U}(i\omega,i\omega',i\Omega)$ from a given impurity fully
irreducible vertex $\delta\Lambda_{\mathrm{imp}}(i\omega,i\omega',i\Omega)$.
Instead of the Dyson equations, one uses the parquet equations which
relate the bare interactions, the fully irreducible vertex and the
fully reducible vertex (see section \ref{subsec:Reminder:-definition-of}):

(i) the lattice fully irreducible vertex $\delta\Lambda_{\mathrm{latt}}$
is approximated by the impurity fully irreducible vertex $\delta\Lambda_{\mathrm{imp}}(i\omega,i\omega',i\Omega)$

(ii) $\delta\Lambda_{\mathrm{latt}}$ is plugged into the lattice
parquet equations (at fixed lattice bare interactions $\hat{U}$)
to get $G_{2,\mathrm{latt}}$, which is summed over the Brillouin
zone to get $G_{2,\mathrm{loc}}(i\omega,i\omega',i\Omega)$

(iii) The impurity parquet equations are inverted to get $\mathcal{U}(i\omega,i\omega',i\Omega)$
from $\delta\Lambda_{\mathrm{imp}}$ and $G_{\mathrm{2,loc}}$.

This construction is remarkable in that the impurity model's bare
interaction $\mathcal{U}_{\sigma_{1}\sigma_{2}\sigma_{3}\sigma_{4}}(i\omega,i\omega',i\Omega)$
is \emph{a priori} different from the lattice's bare interaction $U_{\sigma_{1}\sigma_{2}\sigma_{3}\sigma_{4}}$.
The deviation of $\mathcal{U}_{\sigma_{1}\sigma_{2}\sigma_{3}\sigma_{4}}(i\omega,i\omega',i\Omega)$
from $U_{\sigma_{1}\sigma_{2}\sigma_{3}\sigma_{4}}$, both from the
point of view of the frequency dependence and from the point of view
of the spin structure, is an interesting topic of investigation. We
discuss this in more detail in section \ref{sec:Discussion}.

\subsubsection{Self-consistent loop}

Like DMFT, the equations of section \ref{subsec:A-local-expansion}
may be solved in an iterative way. We propose the following self-consistent
loop:
\begin{enumerate}
\item Start with a given $\mathcal{U}(i\omega,i\omega',\Omega)$ and $\mathcal{G}(i\omega)$
and solve the corresponding impurity model, Eq. (\ref{eq:S_imp_quadrilex}),
for $G_{\mathrm{imp}}$ and $G_{\mathrm{2,imp}}$, from which one
can (through the impurity Dyson and inverse parquet equations) compute
$\Sigma_{\mathrm{imp}}(i\omega)$ and $\delta\Lambda_{\mathrm{imp}}(i\omega,i\omega',i\Omega)$;
\item Use Eq. (\ref{eq:Lambda_tilde_approx}) to get a starting point for
a parquet solver on the lattice. This yields $G_{2,\mathrm{latt}}$; 
\item Use $G_{2}$ (or equivalently $F$) to compute $\Sigma(\mathbf{k},i\omega)$
via the Schwinger-Dyson equation {[}Eq. (\ref{eq:Schwinger_Dyson}){]}\footnote{For stability reasons, one may also compute $\Sigma(\mathbf{k},i\omega)$
and $G(\mathbf{k},i\omega)$ simultaneously with the solution of the
lattice parquet equations (step 2).}, and then $G(\mathbf{k},i\omega)$ via the Dyson equation;
\item Compute the new bath $\mathcal{G}$ and interactions $\mathcal{U}$:

\begin{enumerate}
\item Take the local part of $G(\mathbf{k},i\omega)$ to compute the new
Weiss field as (black part in Fig. \ref{fig:The-quadrilex-loop}):
\begin{equation}
\mathcal{G}^{-1}=G_{\mathrm{loc}}^{-1}+\Sigma_{\mathrm{imp}}\label{eq:Weiss_DMFT}
\end{equation}
\item Take the local part of $G_{2,\mathrm{latt}}$, $G_{2,\mathrm{loc}}$
and use the inverse parquet equations (at fixed $G_{2,\mathrm{loc}}$
and $\delta\Lambda_{\mathrm{imp}}$) to compute $\mathcal{U}(i\omega,i\omega',i\Omega)$
(red part), i.e compute $\Lambda_{\mathrm{imp}}^{'}$ from the inverse
parquet (the prime emphasizes the fact that $\Lambda_{\mathrm{imp}}^{'}$
is obtained after using lattice quantities, as opposed to $\delta\Lambda_{\mathrm{imp}}$
which is computed directly from the impurity model) and get $\mathcal{U}$
as:
\begin{equation}
\mathcal{U}=\Lambda_{\mathrm{imp}}^{'}-\delta\Lambda_{\mathrm{imp}}\label{eq:Weiss_quadrilex}
\end{equation}
\end{enumerate}
\item Go back to step 1. until convergence.
\end{enumerate}
This self-consistent cycle is summarized in Fig. \ref{fig:The-quadrilex-loop}.
The dynamical vertex approximation (D$\Gamma$A in its parquet version)
corresponds to the black parts, namely the computation of $G_{2,\mathrm{latt}}$
from a given impurity vertex $\delta\Lambda_{\mathrm{imp}}(i\omega,i\omega',i\Omega)$.
The computation of the updated impurity bare interaction is shown
in red.

\section{Discussion\label{sec:Discussion}}

\subsection{A dynamical vertex approximation, and beyond}

The local approximation of the lattice vertex, Eq. (\ref{eq:Lambda_tilde_approx}),
is the essence of the dynamical vertex approximation (D$\Gamma$A\cite{Toschi2007,Held2014}).
In fact, D$\Gamma$A in its parquet version can be regarded as a one-shot
realization of the 4PI formalism. 

The main difference of QUADRILEX with D$\Gamma$A is that in D$\Gamma$A,
the vertex $\mathcal{U}$ of the impurity model is not renormalized,
namely in D$\Gamma$A, Eq. (\ref{eq:Weiss_quadrilex}) becomes:
\begin{equation}
\mathcal{U}^{\mathrm{D\Gamma A}}(i\omega,i\omega',i\Omega)=U\label{eq:U_imp_DGA-1}
\end{equation}

As will be seen in the next subsection, this condition is \emph{a
priori} valid only in the atomic limit. As soon as the bandwidth is
finite, deviations of $\mathcal{U}$ with respect to $U$ are likely
to appear.

Interestingly, the natural variable of the QUADRILEX approximation
is the fully irreducible vertex $\Lambda$, which is approximated
locally in the parquet version of the dynamical vertex approximation.
This contrasts with the ladder version of D$\Gamma$A, where the local
approximation is done at the level of the irreducible vertex in one
channel, $\Gamma^{r}$. This vertex derives from another functional,
$\Theta^{r}[G,G_{2}]$ {[}Eq.(\ref{eq:derivative_of_the_log}){]}.
In this variant of D$\Gamma$A, sum rules on the susceptibilities
(or equivalently the asymptotic behavior of the self-energy) are violated
unless some corrections (called ``Moriya corrections'' in the D$\Gamma$A
literature\cite{Katanin2009}) are made. 

Another important consequence of our functional construction is that
the QUADRILEX method can capture the feedback of long-range interactions
into the impurity model. In D$\Gamma$A, this feature only appears
in the solution of the lattice parquet equations (through $U$, which
can in principle have a nonlocal part). One can expect nonlocal physics
to be reflected by a sizable frequency dependence of the local interaction,
$\mathcal{U}$. This frequency dependence can capture some nonlocal
effects of nonlocal interactions such as charge-ordering phenomena,
as has been observed \emph{e.g.} in extended DMFT\cite{Ayral2012,Ayral2013,Huang2014}. 

\subsection{An interpolation between atomic physics and collective-mode physics}

As a local expansion of the 4PI functional, QUADRILEX allows to interpolate
between the atomic limit and the limit of the parquet approximation,
which can describe collective modes. 

Indeed, in the atomic limit, all momentum dependences drop out and
one recovers, by construction,
\begin{equation}
\mathcal{U}^{\mathrm{at}}(i\omega,i\omega',i\Omega)=U\label{eq:U_imp_atomic_limit}
\end{equation}

since the impurity's fully irreducible vertex $\Lambda_{\mathrm{imp}}$
is the exact fully irreducible vertex of the lattice. 

In the weak-coupling regime, $\Lambda_{\mathrm{imp}}$ is roughly
equal to the bare interaction $U$ (the first correction is of order
$U^{4}$), which corresponds to the parquet approximation. Although
the precise form of the impurity's bare interaction $\mathcal{U}_{\mathrm{imp}}(i\omega,i\omega',i\Omega)$
in this limit is difficult to predict \emph{a priori}, one may speculate
that already in this limit a differentiation between the charge and
the spin channel occurs. Such a differentiation has been observed
within the two-particle self-consistent method (TPSC\cite{Vilk1994,Vilk1996,Vilk1997,Tremblay2012}). 

\subsection{Avoiding the parquet: bosonic variables}

\begin{table*}
\bigskip{}

\begin{tabular}{>{\centering}p{0.22\textwidth}>{\centering}p{0.24\textwidth}>{\centering}p{0.15\textwidth}>{\centering}p{0.15\textwidth}>{\centering}p{0.18\textwidth}}
\toprule 
\addlinespace
Degree of irreducibility & Functional & Full correlators & Irred. object & Bare ``bath''\tabularnewline\addlinespace
\midrule
\midrule 
\addlinespace
2PI (DMFT) & $\Phi_{\mathrm{LW}}[G,U]:$ $\Sigma=\frac{\partial\Phi_{\mathrm{LW}}}{\partial G}$ & $G$ & $\Sigma$ & $\mathcal{G}_{i\omega}^{-1}$\tabularnewline\addlinespace
\midrule 
\addlinespace
2PI (EDMFT) & $\Psi[G,W,\lambda]:$ $P=-2\frac{\partial\Psi}{\partial W}$ & $G$, $W$ & $\Sigma$,$P$ & $\mathcal{G}_{i\omega}^{-1}$, $\mathcal{U}_{i\Omega}^{-1}$\tabularnewline\addlinespace
\midrule 
\addlinespace
3PI (TRILEX) & $\mathcal{K}_{3}[G,W,\chi_{3}]:$ $K=-\frac{\partial\mathcal{K}_{3}}{\partial\chi_{3}}$ & \emph{$\chi_{3}$, }$G$, $W$ & $K$ & $\lambda_{i\omega,i\Omega}^{\mathrm{imp}}$, $\mathcal{G}_{i\omega}^{-1}$
, $\mathcal{U}_{i\Omega}^{-1}$\tabularnewline\addlinespace
\midrule 
\addlinespace
4PI (QUADRILEX) & $\mathcal{K}_{4}[G,G_{2}]:$ $\delta\Lambda=-2\frac{\partial\mathcal{K}_{4}}{\partial G_{2}}$ &  $G_{2}$, $G$ & $\delta\Lambda$ & $\mathcal{U}_{i\omega,i\omega',i\Omega}$, $\mathcal{G}_{i\omega}^{-1}$\tabularnewline\addlinespace
\bottomrule
\end{tabular}

\caption{Comparison of the observables for various degrees of irreducibilities.
While 2PI observables are related via Dyson equations, 3PI observables
are simply related by a linear equation, and 4PI observables are related
by parquet equations (see text for definitions).\label{tab:Comparison-of-the}}
\end{table*}

From a technical point of view, the present method has two difficulties:

(i) The solution of parquet equations on the lattice is still a major
computational hurdle despite recent progress\cite{Tam2013,Li2015a}.
This has so far limited the range of applications of parquet-based
methods such as parquet-D$\Gamma$A to very small system sizes.\cite{Valli2014}

(ii) The impurity model Eq. (\ref{eq:S_imp_quadrilex}) features dynamical
interactions with three independent time variables. A straightforward
extension of the existing implementations\cite{Parcollet2014} of
the interaction-expansion continuous-time quantum Monte-Carlo algorithm\cite{Gull2011}
allows to handle this type of problems. It consists in using interaction
vertices depending on four time variables instead of one in the Monte-Carlo
configurations. However, one cannot foretell the severity of the minus
sign problem in the expansion of interaction with three independent
frequencies, which could limit the practical applicability of the
method. We nonetheless note that in some physical regimes, one may
expect the frequency dependence of $\mathcal{U}$ to be restricted
to the bosonic frequency. In this case, the hybridization expansion
solver with dynamical interactions, which is sign-problem-free in
the single-band, density-density case, can be used\cite{Werner2007}. 

An alternative to dealing with the complexity of the four-fermion
interaction is provided by another class of methods also inspired
by DMFT, but relying on the introduction of auxiliary bosonic degrees
of freedom guided by physical insight into the instabilities of the
system. These methods include the extended DMFT method (EDMFT\cite{Sengupta1995,Kajueter1996,Si1996}),
the recently introduced TRILEX method\cite{Ayral2015,Ayral2015c}. 

EDMFT is the straightforward extension of DMFT to electron-boson problems
with fermionic fields (propagator $G$) and bosonic fields (propagator
$W$). In this context, the 2PI functional is denoted as $\Psi[G,W,\lambda]$,
where $\lambda$ is the bare electron-boson coupling. All the fermionic
equations of DMFT can be transposed for bosonic variables, namely
for the bosonic propagator $W$, the bosonic self-energy $P$ and
the bosonic free propagator $W_{0}$. 

The TRILEX method takes EDMFT from the 2PI level (and its functional
$\Psi[G,W,\lambda]$) to the 3PI level and its functional $\mathcal{K}_{3}[G,W,\chi_{3}]$,
where $\chi_{3}$ is now the fully three-point electron-boson correlation
function. Similarly to DMFT and QUADRILEX, it consists in a local
expansion of a functional, which is, in the case of TRILEX, the 3PI
functional $\mathcal{K}_{3}$.

In this approach, the impurity problem has three ``Lagrange multipliers''
or dynamical baths, $\mathcal{G}$, $\mathcal{W}$ and $\lambda^{\mathrm{imp}}$
to satisfy the three constaints on $G_{\mathrm{imp}}$, $W_{\mathrm{imp}}$
and $\chi_{3,\mathrm{imp}}$. The atomic approximation of $\mathcal{K}_{3}$
results in a local approximation of the irreducible vertex
\[
K(\mathbf{k},\mathbf{q},i\omega,i\Omega)=K_{\mathrm{imp}}(i\omega,i\Omega)
\]

Table \ref{tab:Comparison-of-the} summarizes the various methods.
Each method corresponds to the atomic approximation of the nontrivial
part ($\Phi_{\mathrm{LW}}[G,U]$, $\Psi[G,W,\lambda]$, $\mathcal{K}_{3}[G,W,\chi_{3}]$
or $\mathcal{K}_{4}[G,G_{2}]$ respectively) of a functional ($\Gamma_{2}[G,U]$,
$\Gamma_{2}[G,W,\lambda]$, $\Gamma_{3}[G,W,\chi_{3}]$ or $\Gamma_{4}[G,G_{2}]$
respectively). Stationarity of this functional is equivalent to the
fulfillment of ``boldification'' equations (fermionic or bosonic
Dyson equation, simple linear relation (Eq. 18 in Ref. \onlinecite{Ayral2015c}),
or parquet equations, respectively) relating three kinds of objects.
We will call these objects the \emph{``full'' or bold objects} (the
fermionic or bosonic Green's functions $G$ and $W$, connected three-
or four-point functions $\chi_{3}$ and $G_{2}$, respectively), the
\emph{``irreducible'' objects} (the self-energy $\Sigma$, the polarization
$P$, the irreducible three-leg vertex $K$ or fully irreducible four-leg
vertex $\delta\Lambda$) and the \emph{``bare'' objects} ($G_{0}^{-1}$,
$W_{0}^{-1}$, $\lambda$ or $U$ at the lattice level, $\mathcal{G}_{i\omega}^{-1}$,
$\mathcal{U}_{i\Omega}^{-1}$, $\lambda_{i\omega,i\Omega}^{\mathrm{imp}}$
or $\mathcal{U}_{i\omega,i\omega',i\Omega}$ at the impurity level).\footnote{We use the inverse propagators $G_{0}^{-1}$ and $W_{0}^{-1}$ since
these are the objects appearing in the quadratic part of the action.}

In each method, the impurity model is used to compute the irreducible
objects, which are used as an approximation of the corresponding lattice
irreducible object via the atomic approximation of the corresponding
functional. As mentioned before, the variables of this functional
are the full objects, and the atomic approximation imposes that the
local components of these objects coincide with their impurity counterparts.
The knowledge of the full and irreducible objects at the impurity
level allows to find, through the boldification equations, the third
object, namely the bare object which is used as a bath or retarded
interaction in the impurity model.

Other recent methods use bosonic variables to avoid the solution of
parquet equations. A recent example is the dual boson method\cite{Rubtsov2011,VanLoon2014,Stepanov2015}:
it introduces different auxiliary bosonic degrees of freedom, and
relies on the solution of the same impurity model as extended DMFT
(with updated baths in the self-consistent version of the method),
and the subsequent resummation of a selected subclass of self-energy
diagrams built with the lowest-order impurity vertices.

\section{Conclusion and perspectives\label{sec:Conclusion-and-perspectives}}

In this paper, we have used the four-particle irreducible formalism
to build an expansion of the 4PI functional $\mathcal{K}_{4}$ around
the atomic limit. This approximation implies a local approximation
of the fully irreducible vertex, like in the dynamical vertex approximation
(D$\Gamma$A). It maps the lattice model onto an effective local impurity
model with both a dynamical field $\mathcal{G}$ and \emph{dynamical
interactions $\mathcal{U}$ with three independent frequencies} (contrary
to D$\Gamma$A). By construction, this method extrapolates between
the atomic limit at strong coupling and the description of collective
modes by parquet equations at weak coupling.

This functional derivation naturally extends the DMFT idea \textendash{}
a local expansion of the 2PI functional \textendash{} to the 4PI level
and sets up a framework to understand how to generate extensions of
DMFT. In particular, it gives prescriptions as to how to construct
the parameters of the local auxiliary model of DMFT-like approaches.
It is applicable to models with local, but also \emph{nonlocal interactions}
where we expect the dynamical aspect of the interactions to play an
important role.

This functional construction lays the groundwork for more advanced
inquiries about the preservation of conservation laws. One may speculate
that going from 2PI self-consistent (``$\Phi$-derivable'') approximations
to ``$\mathcal{K}_{4}$-derivable'' approximations endows one with
better conservation properties.

The actual implementation of this method is within computational reach
and work is in progress in this direction. On the one hand, the solution
of lattice parquet equations is already part of the parquet-D$\Gamma$A
method and has benefited from recent progress.\cite{Tam2013,Li2015a}
On the other hand, impurity models with dynamical, three-frequency
interactions can be handled by interaction-expansion continuous-time
quantum Monte-Carlo solvers, provided the sign problem is not too
severe. 
\begin{acknowledgments}
We would like to thank A. Toschi and N. Wentzell for a careful reading
of the manuscript, and J.P. Blaizot and S. Andergassen for useful
discussions. This work is supported by the FP7/ERC, under Grant Agreement
No. 278472-MottMetals.
\end{acknowledgments}

\appendix

\section{Properties of the $\zeta$ tensor\label{sec:Properties-of-the-zeta-tensor}}

The tensor $\zeta^{rr'}$ is made of 0's and 1's, and obeys the relation
(to be read as a matrix product in combined $(\alpha\beta)$ indices):
\begin{equation}
\zeta^{rr'}\cdot\zeta^{r'r}=\mathbf{1}\label{eq:closure_zeta}
\end{equation}

Indeed, for any $\hat{C}$,
\[
\hat{C}_{r,\bar{\gamma}\bar{\delta}}=\zeta_{\bar{\gamma}\bar{\delta},\alpha\beta}^{rr'}\hat{C}_{r',\alpha\beta}=\zeta_{\bar{\gamma}\bar{\delta},\alpha\beta}^{rr'}\zeta_{\alpha\beta,\gamma\delta}^{r'r}\hat{C}_{r,\gamma\delta}
\]
and therefore $\zeta_{\bar{\gamma}\bar{\delta},\alpha\beta}^{rr'}\zeta_{\alpha\beta,\gamma\delta}^{r'r}=\delta_{\bar{\gamma},\gamma}\delta_{\bar{\delta}\delta}$.

\section{Relation between $G_{2}$ and $F$ in channel notation\label{sec:Relation-between-}}

In this section, we prove Eq. (\ref{eq:G2_XFX}). We have:
\begin{align*}
\hat{\chi}_{\mathrm{ph},\alpha\delta}^{\mathrm{ph}}\hat{F}_{\mathrm{ph},\delta\gamma}\hat{\chi}_{\mathrm{ph},\gamma\beta}^{\mathrm{ph}} & =\hat{\chi}_{\mathrm{ph},\bar{a}b,a\bar{b}}^{\mathrm{ph}}\hat{F}_{\mathrm{ph},a\bar{b},\bar{v}v}\hat{\chi}_{\mathrm{ph},\bar{v}v,u\bar{u}}^{\mathrm{ph}}\\
 & =G_{a\bar{a}}G_{b\bar{b}}F_{a\bar{v}v\bar{b}}G_{u\bar{v}}G_{v\bar{u}}\\
 & =G_{2,\bar{a}u\bar{u}b}\\
 & =\hat{G}_{2,\mathrm{ph},\bar{a}b,u\bar{u}}\\
 & =\hat{G}_{2,\mathrm{ph},\alpha\beta}
\end{align*}

\begin{align*}
\hat{\chi}_{\mathrm{\overline{ph}},\alpha\delta}^{\mathrm{\overline{ph}}}\hat{F}_{\mathrm{\mathrm{\overline{ph}}},\delta\gamma}\hat{\chi}_{\mathrm{\overline{ph}},\gamma\beta}^{\mathrm{\mathrm{\overline{ph}}}} & =\hat{\chi}_{\mathrm{\overline{ph}},\bar{u}u,\bar{v}v}^{\mathrm{\overline{ph}}}\hat{F}_{\mathrm{\overline{ph}},\bar{v}v,\bar{a}a}\hat{\chi}_{\mathrm{\overline{ph}},\bar{a}a,\bar{b}b}^{\mathrm{\overline{ph}}}\\
 & =G_{u\bar{v}}G_{v\bar{u}}F_{v\bar{v}a\bar{a}}G_{a\bar{b}}G_{b\bar{a}}\\
 & =G_{2,\bar{u}u\bar{b}b}\\
 & =G_{2,\bar{b}b\bar{u}u}\\
 & =\hat{G}_{2,\overline{\mathrm{ph}},\bar{u}u,\bar{b}b}\\
 & =\hat{G}_{2,\overline{\mathrm{ph}},\alpha\beta}
\end{align*}

and

\begin{align*}
\hat{\chi}_{\mathrm{pp},\alpha\delta}^{\mathrm{pp}}\hat{F}_{\mathrm{pp},\delta\gamma}\hat{\chi}_{\mathrm{pp},\gamma\beta}^{\mathrm{\mathrm{pp}}} & =\hat{\chi}_{\mathrm{pp},\bar{u}\bar{v},uv}^{\mathrm{pp}}\hat{F}_{\mathrm{pp},uv,\bar{a}\bar{b}}\hat{\chi}_{\mathrm{pp},\bar{a}\bar{b},ab}^{\mathrm{pp}}\\
 & =G_{v\bar{u}}G_{u\bar{v}}F_{u\bar{a}v\bar{b}}G_{a\bar{b}}G_{b\bar{a}}\\
 & =G_{2,\bar{v}b\bar{u}a}\\
 & =G_{2,\bar{u}a\bar{v}b}\\
 & =\hat{G}_{2,\mathrm{pp},\bar{u}\bar{v},ab}\\
 & =\hat{G}_{2,\mathrm{pp},\alpha\beta}
\end{align*}

We have used the crossing symmetry 
\begin{equation}
G_{2,\bar{u}u\bar{v}v}=G_{2,\bar{v}v\bar{u}u}\label{eq:crossing_sym}
\end{equation}
for the last two equalities. 

\section{Consistency of the Self-Energy in the 4PI formalism\label{sec:Self-Energy}}

Here, we show that $\Sigma$ given by the Schwinger-Dyson equation
is identical to the derivative of the Luttinger-Ward functional with
respect to $G$ (Eq. (\ref{eq:Sigma_LW})) based on the homogeneity
properties of the 4PI functional.

We first express $\Phi_{\mathrm{LW}}$ as a function of $G$, $U$
and $G_{2}$ using Eq. (\ref{eq:Kappa4_def}):
\begin{equation}
\tilde{\Phi}_{\mathrm{LW}}[G,U,G_{2}]\equiv R[G,G_{2}]-\frac{1}{2}\hat{U}_{r,\alpha\beta}\hat{G}_{2,r,\beta\alpha}^{\mathrm{nc}}\label{eq:Kappa4_def-1}
\end{equation}

with
\begin{eqnarray}
 &  & R[G,G_{2}]\equiv\mathcal{K}_{4}[G,G_{2}]+\frac{1}{2}\sum_{r}\Theta^{r}[G,G_{2}]\nonumber \\
 &  & -\frac{1}{2}\left(\hat{\chi}_{r,\alpha\gamma}^{r}\right)^{-1}\hat{G}_{2,\gamma\delta}\left(\hat{\chi}_{r,\delta\beta}^{r}\right)^{-1}\hat{G}_{2,r,\beta\alpha}\label{eq:R_def}
\end{eqnarray}

Then, following (\ref{eq:Sigma_LW}), $\Sigma$ is given by:
\[
\Sigma_{\bar{u}v}=\frac{\partial\tilde{\Phi}_{\mathrm{LW}}}{\partial G_{v\bar{u}}}\Bigg|_{U,G_{2}}+\frac{\partial\tilde{\Phi}_{\mathrm{LW}}}{\partial\hat{G}_{2,r,\alpha\beta}}\Bigg|_{U,G}\frac{\partial\hat{G}_{2,r,\alpha\beta}}{\partial G_{v\bar{u}}}
\]

The second term evaluates to zero. Indeed, we can first, using Eqs
(\ref{eq:Gamma4_def}) and (\ref{eq:Gamma2_decomp}), see that
\begin{equation}
R[G,G_{2}]=\Gamma_{4}[G,G_{2}]-\Gamma_{2,0}[G]\label{eq:R_prop}
\end{equation}

Then, using Eq. (\ref{eq:R_prop}), the fact that $\Gamma_{2,0}$
does not depend on $G_{2}$ and Eq (\ref{eq:reciprocity_Gamma4}),
we obtain:

\begin{align}
\frac{\partial\tilde{\Phi}_{\mathrm{LW}}}{\partial\hat{G}_{2,r,\alpha\beta}}\Bigg|_{U,G} & =\frac{\partial\Gamma_{4}}{\partial\hat{G}_{2,r,\alpha\beta}}\Bigg|_{U,G}-\frac{1}{2}\hat{U}_{r,\beta\alpha}=0\label{eq:vanishing_G2_der}
\end{align}

Hence:
\begin{align}
\Sigma_{\bar{u}v} & =\frac{\partial R}{\partial G_{v\bar{u}}}\Bigg|_{U,G_{2}}+\Sigma_{\bar{u}v}^{\mathrm{HF}}\label{eq:Sigma_intermediate-1}
\end{align}

where we have defined:
\begin{align*}
\Sigma_{\bar{u}v}^{\mathrm{HF}} & \equiv-\frac{1}{2}\frac{\partial}{\partial G_{v\bar{u}}}\left[U_{a\bar{a}b\bar{b}}\left(-G_{a\bar{a}}G_{b\bar{b}}+G_{a\bar{b}}G_{b\bar{a}}\right)\right]\\
 & =\frac{1}{2}U_{a\bar{a}b\bar{b}}\Big(\delta_{av}\delta_{\bar{u}\bar{a}}G_{b\bar{b}}+G_{a\bar{a}}\delta_{vb}\delta_{\bar{u}\bar{b}}\\
 & \;\;-\delta_{va}\delta_{\bar{u}\bar{b}}G_{b\bar{a}}-G_{a\bar{b}}\delta_{vb}\delta_{\bar{u}\bar{a}}\Big)\\
 & =\frac{1}{2}\Big(U_{v\bar{u}b\bar{b}}G_{b\bar{b}}+U_{a\bar{a}v\bar{u}}G_{a\bar{a}}\\
 & \;\;-U_{v\bar{a}b\bar{u}}G_{b\bar{a}}-U_{a\bar{u}v\bar{b}}G_{a\bar{b}}\Big)\\
 & =U_{v\bar{u}a\bar{a}}G_{a\bar{a}}-U_{v\bar{a}a\bar{u}}G_{a\bar{a}}
\end{align*}

To obtain the last line, we have used the crossing symmetry (\ref{eq:crossing_sym})
and relabelled the indices. This is the Hartree-Fock term.

As for the first term of (\ref{eq:Sigma_intermediate-1}), it can
be rewritten using the homogeneity properties of $R$. Let us first
show that $R$ is the sum of homogeneous functions of the function
$Y^{r}$, namely
\begin{equation}
R[G,G_{2}]=\sum_{r}\tilde{R}^{r}[\hat{Y}^{r}]\label{eq:R_homogeneity}
\end{equation}

with:
\begin{equation}
\hat{Y}^{r}\equiv\left(\hat{\chi}^{r}\right)^{-1}\hat{G}_{2}\label{eq:Y_def}
\end{equation}

\begin{figure}
\begin{centering}
\includegraphics[width=0.5\columnwidth]{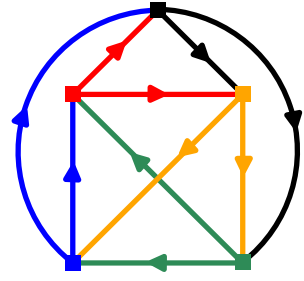}
\par\end{centering}
\caption{(color online) Homogeneity of the lowest-order diagram of $\mathcal{K}_{4}$
in $\hat{Y}^{\mathrm{pp}}$. Each colored piece stands for $\hat{Y}^{\mathrm{pp}}=(\hat{\chi}^{\mathrm{pp}})^{-1}\hat{G}_{2}=\hat{F}\hat{\chi}^{\mathrm{pp}}$.
\label{fig:Homogeneity-of-the}}
\end{figure}

This property follows from the homogeneity of the three terms in the
definition of $R$, Eq. (\ref{eq:R_def}). $\mathcal{K}_{4}$ is homogeneous
with respect to $Y^{r}$ in the three channels:
\begin{equation}
\mathcal{K}_{4}[G,G_{2}]=\tilde{\mathcal{K}}_{4}^{\mathrm{ph}}[Y^{\mathrm{ph}}]=\tilde{\mathcal{K}}_{4}^{\overline{\mathrm{ph}}}[Y^{\overline{\mathrm{ph}}}]=\tilde{\mathcal{K}}_{4}^{\mathrm{pp}}[Y^{\mathrm{pp}}]\label{eq:Kappa_homog}
\end{equation}

The homogeneity of the lowest diagram of $\mathcal{K}_{4}$ with respect
to $Y^{\mathrm{pp}}$ is illustrated in Figure \ref{fig:Homogeneity-of-the}.
The homogeneity of $\Theta^{r}$ with respect to $Y^{r}$ follows
from its definition {[}Eq.(\ref{eq:def_Theta}){]} and the cyclity
of the trace: one can easily check that 
\begin{equation}
\Theta^{r}[G,G_{2}]=\tilde{\Theta}^{r}[\hat{G}_{2}(\hat{\chi}^{r})^{-1}]=\tilde{\Theta}^{r}[(\hat{\chi}^{r})^{-1}\hat{G}_{2}]\label{eq:Theta_homog}
\end{equation}
The last term in Eq. (\ref{eq:R_def}) is obviously homogeneous in
$Y^{r}$ for all $r$'s.

We now use Eq.(\ref{eq:R_homogeneity}) to decompose:
\[
\frac{\partial R}{\partial\hat{G}_{v\bar{u}}}\Bigg|_{U,G}=\sum_{r}\frac{\partial\tilde{R}^{r}}{\partial\hat{G}_{v\bar{u}}}\Bigg|_{U,G}
\]

For a given $r$, we first use the chain rule to write:

\begin{equation}
\frac{\partial\tilde{R}^{r}}{\partial\hat{G}_{v\bar{u}}}\Bigg|_{U,G_{2}}=\frac{\partial\tilde{R}^{r}}{\partial\left(\hat{\chi}^{r}\right)_{r,\mu\nu}^{-1}}\frac{\partial\left(\hat{\chi}^{r}\right)_{r,\mu\nu}^{-1}}{\partial\hat{G}_{v\bar{u}}}\label{eq:dK4_dG_interm}
\end{equation}

The first factor evaluates to: 
\begin{align*}
\frac{\partial\tilde{R}^{r}}{\partial\left(\hat{\chi}^{r}\right)_{r,\alpha\beta}^{-1}}\Bigg|_{G_{2}} & =\frac{\partial\tilde{R}^{r}}{\partial\hat{Y}_{r',\mu\nu}^{r}}\frac{\partial\hat{Y}_{r',\mu\nu}^{r}}{\partial\left(\hat{\chi}^{r}\right)_{r,\alpha\beta}^{-1}}=\hat{G}_{2,r',\beta\nu}\frac{\partial\tilde{R}^{r}}{\partial\hat{Y}_{r',\alpha\nu}^{r}}
\end{align*}

In the right-hand side, the first factor evaluates to: 
\begin{equation}
\frac{\partial\tilde{R}^{r}}{\partial\hat{Y}_{r,\gamma\beta}^{r}}=\frac{\partial\tilde{R}^{r}}{\partial\hat{G}_{2,r,\alpha\beta}}\Bigg|_{G}\hat{\chi}_{r,\alpha\gamma}^{r}\label{eq:partial_Y_vs_partial_G2}
\end{equation}

Indeed,
\begin{align*}
\frac{\partial\tilde{R}^{r}}{\partial\hat{G}_{2,r,\alpha\beta}}\Bigg|_{G} & =\frac{\partial\tilde{R}^{r}}{\partial\hat{Y}_{r,\mu\nu}^{r}}\frac{\partial\hat{Y}_{r,\mu\nu}^{r}}{\partial\hat{G}_{2,r,\alpha\beta}}=\left(\hat{\chi}^{r}\right)_{r,\mu\alpha}^{-1}\frac{\partial\tilde{R}^{r}}{\partial\hat{Y}_{r,\mu\beta}^{r}}
\end{align*}

Thus, Eq. (\ref{eq:dK4_dG_interm}) becomes:

\begin{align}
\frac{\partial\tilde{R}^{r}}{\partial\hat{G}_{v\bar{u}}}\Bigg|_{U,G} & =\hat{G}_{2,r,\nu\gamma}\frac{\partial\tilde{R}^{r}}{\partial\hat{G}_{2,r,\alpha\gamma}}\Bigg|_{G}\hat{\chi}_{r,\alpha\mu}^{r}\frac{\partial\left(\hat{\chi}^{r}\right)_{r,\mu\nu}^{-1}}{\partial G_{v\bar{u}}}\label{eq:dR_dG_interm}
\end{align}

Let us define
\begin{equation}
\hat{U}_{r,\gamma\alpha}^{r}\equiv2\frac{\partial\tilde{R}^{r}}{\partial\hat{G}_{2,r,\alpha\gamma}}\Bigg|_{G}\label{eq:U_r_def}
\end{equation}

Plugging Eq.(\ref{eq:dR_dG_interm}) into Eq.(\ref{eq:Sigma_intermediate-1}),
we thus end up with {[}using Eq. (\ref{eq:G2_XFX}) and the cyclicity
of the trace{]}:
\begin{align}
\Sigma_{\bar{u}v}-\Sigma_{\bar{u}v}^{\mathrm{HF}} & =\sum_{r}\mathrm{Tr}\left[\hat{G}_{2}\frac{1}{2}\hat{U}^{r}\hat{\chi}^{r}\frac{\partial\left(\hat{\chi}^{r}\right)^{-1}}{\partial G_{v\bar{u}}}\right]\nonumber \\
 & =\frac{1}{2}\sum_{r}\mathrm{Tr}\left[\hat{F}\hat{\chi}^{r}\hat{U}^{r}\hat{\chi}^{r}\frac{\partial\left(\hat{\chi}^{r}\right)^{-1}}{\partial G_{v\bar{u}}}\hat{\chi}^{r}\right]\nonumber \\
 & =-\frac{1}{2}\sum_{r}\mathrm{Tr}\left[\hat{F}\hat{\chi}^{r}\hat{U}^{r}\frac{\partial\hat{\chi}^{r}}{\partial G_{v\bar{u}}}\right]\label{eq:Sigma_as_trace_FU}
\end{align}

In addition to being homogeneous with respect to $\hat{Y}^{r}$ {[}Eq.
(\ref{eq:R_homogeneity}){]}, $R$ is homogeneous with respect to
its transpose $\left(\hat{Y}^{r}\right)^{T}$. With this property,
the same steps as above lead to the expression

\begin{equation}
\Sigma_{\bar{u}v}-\Sigma_{\bar{u}v}^{\mathrm{HF}}=-\frac{1}{2}\sum_{r}\mathrm{Tr}\left[\hat{U}^{r}\hat{\chi}^{r}\hat{F}\frac{\partial\hat{\chi}^{r}}{\partial G_{v\bar{u}}}\right]\label{eq:Sigma_as_trace_UF}
\end{equation}

Expressing $\Sigma-\Sigma^{\mathrm{HF}}$ as the half sum of the right-hand
sides Eqs (\ref{eq:Sigma_as_trace_FU}) and (\ref{eq:Sigma_as_trace_UF})
and expanding the latter using the change of notation defined in Eqs
(\ref{eq:C_ph}-\ref{eq:C_ph_bar}-\ref{eq:C_pp}), (\ref{eq:V_ph}-\ref{eq:V_ph_bar}-\ref{eq:V_pp}),
we find that:

\begin{align}
\Sigma_{\bar{u}v}-\Sigma_{\bar{u}v}^{\mathrm{HF}} & =-\frac{1}{2}F_{b\bar{a}c\bar{u}}G_{a\bar{a}}G_{c\bar{c}}G_{b\bar{b}}\left(\sum_{r}U_{a\bar{b}v\bar{c}}^{r}\right)\nonumber \\
 & \;\;-\frac{1}{2}\left(\sum_{r}U_{b\bar{a}c\bar{u}}^{r}\right)G_{a\bar{a}}G_{c\bar{c}}G_{b\bar{b}}F_{a\bar{b}v\bar{c}}\nonumber \\
 & =-\frac{1}{2}F_{b\bar{a}c\bar{u}}G_{a\bar{a}}G_{c\bar{c}}G_{b\bar{b}}U_{a\bar{b}v\bar{c}}\nonumber \\
 & \;\;-\frac{1}{2}U_{b\bar{a}c\bar{u}}G_{a\bar{a}}G_{c\bar{c}}G_{b\bar{b}}F_{a\bar{b}v\bar{c}}\label{eq:Sigma_SD}
\end{align}

To obtain the last expression, we have used the identity
\begin{equation}
\sum_{r}\hat{U}^{r}=\hat{U}\label{eq:sum_U_r}
\end{equation}

which follows from Eq.(\ref{eq:vanishing_G2_der}).

Eq. (\ref{eq:Sigma_SD}) is nothing but the Schwinger-Dyson equation
{[}Eq.(\ref{eq:Schwinger_Dyson}){]} for the self-energy (the two
terms are equal, see \emph{e.g.} Ref. \onlinecite{VanLeeuwen2006})

\bibliographystyle{apsrev4-1}
\bibliography{library}

\end{document}